\documentstyle[12pt,fleqn,epsf,epsfig]{article}
\begin{document}

\title{Symbiotic Sympatric Speciation: 
Compliance with Interaction-driven Phenotype 
Differentiation from a Single Genotype}

\author{
        Kunihiko Kaneko\\
        {\small \sl Department of Pure and Applied Sciences}\\
        {\small \sl University of Tokyo, Komaba, Meguro-ku, Tokyo 153,
JAPAN}\\
}

\date{}
\vspace{.1in}
\maketitle

\vspace{1in}


\vspace{.2in}
--------------

e-mail: kaneko@complex.c.u-tokyo.ac.jp

\pagebreak
\vspace{.3in}

\begin{abstract}
A mechanism of sympatric speciation is presented based on the
interaction-induced developmental plasticity  of phenotypes. First, phenotypes of individuals with
identical genotypes split into a few groups, according to instability
in the developmental dynamics that are triggered with the competitive interaction
among individuals.  Then, through mutational change of genes, the
phenotypic differences are fixed to genes, until the groups
are completely separated in genes as well as phenotypes.
It is also demonstrated that the proposed
theory leads to hybrid sterility under sexual recombination, and thus
speciation is completed in the sense of reproductive isolation.
As a result of this post-mating isolation, the mating preference evolves later.
When there are two alleles, the correlation between alleles is formed, to consolidate the speciation.
When individuals are located in space, different species are later segregated
spatially, implying that the speciation so far regarded to be allopatric
may be a result of sympatric speciation.
Relationships with previous theories, frequency-dependent selection, reinforcement,
Baldwin's effect, phenotypic plasticity, and resource competition are briefly discussed.
Relevance of the results to natural evolution are
discussed, including punctuated equilibrium, incomplete penetrance in mutants, and
the change in flexibility in genotype-phenotype correspondence.
Finally, it is discussed how our theory is confirmed 
both in field and in laboratory (in an experiment with the use of {\sl E coli.}).
\end{abstract}

\pagebreak
\section{Introduction}

In spite of progress in the understanding in evolution ever since Darwin(1869),
the speciation is not yet fully understood.
In the recent book, Maynard-Smith and Szathmary(1995) wrote that
{\sl we are not aware of any explicit
model demonstrating the instability of a sexual continuum}.

To discuss the problem of speciation,
let us start from reviewing  basic 
standpoints  in evolution theory, although it might look too elementary here.

\begin{itemize}

\item

(i) Existence of genotype and phenotype

\item

(ii) Fitness for reproduction is given as a function of the phenotype and
the environment.  The ``environment" can include
interaction with other individuals. In other words,
the reproduction rate of an individual is a function
of its phenotype, and environment, i.e., $F($phenotype, environment).

\item

(iii) Only the genotype is transferred to the next generation
(Weissman doctrine)

\item

(iv) There is flow only from genotype to phenotype
(the central dogma of the molecular biology).  For example,
through the developmental process the phenotype is determined
depending on the genotype. Now, the process is summarized as
{\sl Genotype $\rightarrow$ Development $\rightarrow$ Phenotype}. 

\end{itemize}

Here we adopt these standard assumptions.
( Although the assumption (iii) may not be valid for some cases
known as epigenetic inheritance,
we accept the assumption here, since the relevance of
epigenetic inheritance to
evolution is still controversial, and the theory to be proposed
is valid in the presence of epigenetic inheritance, but
does not require it.)

In the standard evolutionary genetics, the assumption (iv) is
further replaced by a stronger one, i.e., (iv') ``phenotype is
a {\em single} valued function of genotype''.  If this were always true,
we could replace $F($phenotype,environment) in (i) by
$F(f$(genotype),environment) and then we could discuss the evolutionary
process in terms of the population dynamics only of genotypes
(and environment).  This is the basic standpoint in
population genetics.

Indeed, this reduction to genes is valid for gradual evolution.
It is also supported by the following mathematical argument.
The change of genotype is slower in time scale than that of phenotype.
As is known, variables with slower time scale act as a control parameter
to faster ones, if the time scale separation is large enough
(and if the dynamics in the fast time scale do not have such instability
that leads to bifurcation).

Still, explanation of the speciation, especially sympatric speciation,
is not so easy following this standard evolutionary genetics.
If slight genetic change leads to slight phenotype change,
then individuals arising from mutation from the same genetic group
differ only slightly according to this picture.
Then, these individuals  compete each other for the same niche.
Unless the phenotype in concern is neutral, it is 
generally difficult that two (or more) groups coexist.  Those with a
higher fitness would survive.
One possible way to get out of this difficulty is to 
assume that two groups are 
`effectively' isolated, so that they do not compete.  
Some candidates for such isolation are searched. The most
well-known example is spatial segregation, known as allopatric speciation.
Since we are here interested in sympatric speciation, this
solution cannot be adopted here\footnote{As will be discussed later,
sympatric speciation can bring about allopatric one, but not the other
way round.} Furthermore, there are direct evidences that
sympatric speciation really occurred in the evolution, for example
in the speciation of cichlid in some lakes(Schiliewen et al. 1994).

As another candidate for separation mating preference is discussed
(Maynard-Smith 1966,  Felsenstein 1981, Grant 1981, Doebeli 1996,
Howard and Barlocher eds. 1998).  Recently, there have appeared some
models showing the instability of sexual continuum, without assuming the
existence of discrete groups in the beginning.
Probably, the argument based on the runaway
is most persuasive (Lande 1981, Turner and Burrows 1995,Howard and Barlocher eds. 1998).
Even though
two groups coexist at the same spatial location, they can be genetically separated
if two groups do not mate each other.  Hence, the mating preference is proposed
as a mechanism for sympatric speciation.  However, in this theory, why there is 
such mating preference itself is not answered.  Accordingly, it is not self-contained
as a theory.

Another recent proposal is the introduction of (almost) neutral fitness landscape and
exclusion of individuals with similar phenotypes
(Dieckmann and Doebeli, 1999, Kondrashov and Kondrashov,1999 Kawata and Yoshimura 2001).
For example,
Dieckmann and Doebeli[1999]  have succeeded in showing that two groups
are formed and coexist, to avoid the competition among organisms with similar
phenotypes, assuming a rather flat fitness-landscape.
This provides one explanation and can be relevant to some sympatric speciation.
However, it is not so clear how the phenotype that is not so important as 
a fitness works strongly as a factor for exclusion for a closer value.
Furthermore, we are more interested in the differentiation of phenotypes
that are functionally different and not neutral.  

So far, in these studies, the interaction between individuals lead to
competition for their survival.  Difficulty in stable sympatric speciation without mating preference
lies in the lack of a known
clear mechanism how two groups, which have just started to be separated, 
coexist in the presence of mutual interaction. Of course, if the two groups 
were in a symbiotic state, the coexistence
could help the survival of each.  However, the two groups 
have little difference in genotype in the beginning of speciation process, according to the
assumption (iv)'.  Then,
it would be quite difficult to imagine such a `symbiotic' mechanism.  

Now, the problem we address here is as follows:
If we do not assume (iv)' ( but by assuming (i)-(iv).),
is there any mechanism
that two groups mutually require each other for the survival
in the beginning of the separation of the two groups, 
In the present paper we propose such mechanism, and provide
a sympatric speciation scenario robust against fluctuations.

Note that the above difficulty  comes from the assumption that the phenotype is
a single-valued function of genotype.
Is this single-valued-ness always true?
To address this question, we reconsider the G-P relationship.
Indeed, there are three reasons that we doubt this single-valued-ness.

First, Yomo  and his colleagues have
reported that specific mutants of {\it E. coli}
show (at least) two distinct types of enzyme activity, although they
have identical genes(Ko et al., 1994).  These different types
coexist in an unstructured environment of a chemostat (Ko et al. 1994), and
this coexistence is not due to spatial localization.  Coexistence of
each type is supported by each other.  Indeed, when one type of {\it E. coli} is
removed externally, the remained type starts differentiation again
to recover the coexistence of the two types.
The experiment demonstrates
that the enzyme activity of these {\it E. coli}
are differentiated into two (or more) groups, due to the interaction with
each other, even though they have identical genes.
Here spatial factor is not important, since this experiment is carried out
in a well stirred chemostat.

Second, some organisms are known to show various phenotype from a single genotype.
This phenomenon is often related to malfunctions of a mutant
(Holmes 1979), and is called
as low or incomplete penetrance(Opitz 1981). 

Third, a theoretical  mechanism of phenotypic
diversification has already been proposed as
the isologous diversification for cell
differentiation(Kaneko and Yomo, 1994,1997,1999; Furusawa and Kaneko 1998).
The theory states that phenotypic diversity will
arise from a single genotype and develop dynamically through intracellular
complexity and intercellular connection.
When organisms with
plastic developmental dynamics interact with each other,
the dynamics of each unit can be stabilized by forming distinct groups with
differentiated states in the pheno-space.  Here the two differentiated
groups are necessary to stabilize each of the dynamics.  Otherwise, the developmental
process is unstable, and through the interaction the two types are formed again,
when there is a sufficient number of units.  This theoretical mechanism
is demonstrated by several models and is shown to be a general consequence of
coupled dynamical systems.

The isologous diversification theory shows that there can be
developmental `flexibility', in which different phenotypes arise
from identical gene sets, as in the incomplete penetrance aforementioned.
Now we have to study how this theory is relevant to evolution.
Indeed, the question how developmental process and evolution are related
has been addressed over decades (Maynard-Smith et al.  1985).
We consider correspondence between genotype and phenotype seriously, by
introducing a developmental process with which
a given initial condition is lead to some phenotype according to a given genotype.
`Development' here  means a dynamic process from an initial state
to a matured state through rules associated with genes.
(In this sense, it is not necessarily restricted to multicellular organisms.)

\section{Model}

To consider the evolution with developmental dynamics,
it is appropriate to represent  phenotype by
a set of state variables. For example,
each individual $i$ has variables
$(X^1_t (i)$,$X^2_t(i)$, $\cdots ,X^k_t(i))$, which defines the phenotype.
This set of variables can be regarded as concentrations of
chemicals, rates of metabolic processes, or some
quantity corresponding to a higher function characterizing the behavior of
the organism.  The state is not fixed in time,
but develops from the initial state at birth to a matured state when the
organism is ready to produce its offspring.
The dynamics of the state variables $(X^1_t (i)$,$X^2_t(i)$, $\cdots ,X^k_t(i))$
is given by a set of  equations with some parameters.

Genes, since they are nothing but information expressed on DNA, could in
principle be included in the set of variables. However, according to the
central dogma of molecular biology (requisite (iv)), the gene has a
special role among such variables. Genes can affect phenotypes,
the set of variables, but the
phenotypes cannot change the code of genes. During the life cycle, changes
in genes are negligible compared with those of the phenotypic variables
they control.
In terms of dynamical systems, the set corresponding to genes
can be represented by parameters $\{g^1(i),g^2(i), \cdots g^m(i) \}$
that govern the dynamics of phenotypes, since
the parameters in an equation are not changed through the developmental process,
while the parameters control the dynamics of phenotypic variables.
Accordingly, we represent the genotype by a set of parameters.
Only when an individual organism is reproduced, this set of
parameters changes slightly by mutation.
For example, when $\{X^{\ell}_t (j)\}$ represents the concentrations
of metabolic chemicals, $\{g^1(i),g^2(i), \cdots g^m(i) \}$ is
the catalytic activity of enzymes that controls the corresponding
chemical reaction.

Now, our model is set up as follows:

(1) Dynamical change of states giving a phenotype:

The temporal evolution of the state variables $(X^1_t (i)$,$X^2_t(i)$, $\cdots ,X^k_t(i))$
is given by a set of deterministic equations,
which are described by the state of the individual,
and parameters $\{g^1(i),g^2(i), \cdots g^m(i) \}$ (genotype),
and the interaction with other individuals.
This temporal evolution of the state consists of internal dynamics 
and interaction.

(1-1)The internal dynamics (say metabolic process in an organism)
are represented by the equation governed only of
$(X^1_t (i)$,$X^2_t(i)$, $\cdots ,X^k_t(i))$ ( without dependence on
$\{(X^{\ell}_t (j)\}$ ($j \neq i$)), and are controlled by the
parameter sets  $\{g^1(i),g^2(i), \cdots g^m(i) \}$ .

(1-2) Interaction between the individuals:  The interaction
is given through
the set of variables $(X^1_t (i),X^2_t(i),\cdots ,X^k_t(i))$.
For example, we consider such interaction form that
the individuals interact with all others through competition for
some `resources'.
The resources are taken by all the individuals, giving competition
among all the individuals.   Since we are interested in
sympatric speciation, we take this extreme all-to-all interaction,
by taking a well stirred soup of resources,
without including any spatially localized interaction.
 
(2) Reproduction  and Death:
Each individual gives offspring (or splits into two)
when a given `maturity condition' for growth is satisfied.  This condition is given by
a set of  variables $(X^1_t (i),X^2_t(i),\cdots ,X^k_t(i))$.
For example, if $(X^1_t (i),X^2_t(i),\cdots ,X^k_t(i))$ represents
cyclic process corresponds to a metabolic,
genetic or other process that is required for reproduction, we assume
that the unit replicates when the accumulated number of cyclic processes goes
beyond some threshold.  

(3) Mutation:
When each organism reproduces, the  set of parameters $\{ g^j(i) \}$
changes slightly by mutation, by adding a  random number
with a small amplitude  $\delta$, corresponding to the mutation rate.
The values of variables  $(X^1_t (i),X^2_t(i),\cdots ,X^k_t(i))$  are
not transferred but are reset to initial conditions.
(If one wants to include some factor of epigenetic inheritance
one could assumed that some of the values of state variables are 
transferred.  Indeed we have carried out this simulation also,
but the results to be discussed are not altered (or confirmed more strongly).

(4) Competition:
To introduce competition for survival,
death is included both by random removal of organisms at some rate
as well as by a given death condition based on their state.

For a specific example, see Appendix.

\section{Sympatric Speciation observed}

From several simulations satisfying the condition of the model in \S 2,  we have
obtained a scenario for a sympatric speciation process(Kaneko and Yomo 2000,2002)
The speciation process we observed
is schematically shown in Fig.1, where the change of 
the correspondence between
a phenotypic variable (``P") and a genotypic parameter (``G")  is plotted at every
reproduction event.
This scenario is summarized as follows.

In the beginning, there is a single species, with
one-to-one correspondence between phenotype and genotype.  Here,
there are little genetic and phenotypic diversity that are continuously
distributed.(see Fig.1a).  
We assume that the isologous diversification starts
due to developmental plasticity with interaction,
when the number of these organisms
increase.  Indeed, the existence of such phenotypic differentiation
is supported by isologous diversification,, and
is supported by several numerical experiments.
This gives the following stage-I.

{\bf Stage-I: Interaction-induced phenotypic differentiation}

When there are many individuals interacting for finite resources,
the phenotypic dynamics start to be differentiated even though
the genotypes are identical or differ only slightly.
Phenotypic variables split into two (or more) types (see Fig.1b).
This interaction-induced
differentiation is an outcome of the mechanism aforementioned.
Slight phenotypic difference between individuals is amplified
by the internal dynamics, while
through the interaction between organisms, the difference of
phenotypic dynamics tends to be clustered into two (or more) types.
Here  the two distinct phenotype groups (brought about by interaction) are
called `upper' and `lower' groups, tentatively.

This differentiation is brought about, since the population consisting of
individuals taking identical phenotypes
is destabilized by the interaction.
Such instability is for example, caused by
the increase of population or decrease of resources, leading to
strong competition.  Of course, if the phenotype  $X_t^j(i)$ 
at a matured state is rigidly determined by  developmental dynamics,
such differentiation does not occur.  Only the assumption we make
in the present theory is that there exists such developmental plasticity
in the internal dynamics, when the interaction is strong.
Recall again that this assumption is theoretically supported.

Note that the difference is fixed at this stage neither
at the genetic nor phenotypic level.
After reproduction, an individual's phenotype can switch to another type.

{\bf Stage-II: Amplification of the difference through
genotype-phenotype relationship}

At the second stage the
difference between the two groups is amplified
both at the genotypic and at the phenotypic level.  This is realized by a
kind of positive feedback process between the  change of
geno- and pheno-types.

First the genetic parameter(s) separate as a result of the phenotypic change.
This occurs if the parameter dependence of the growth rate is different
between the two phenotypes.  Generally, there are one or several
parameter(s) $g^{\ell}$, such that the growth rate increases
with $g^{\ell}$ for the upper group and decreases for the lower group
(or the other way round) (see Fig.1c and Fig.2).

Certainly, such a parameter dependence is not exceptional.
As a simple illustration, assume that the use of metabolic processes
is different between the two phenotypic groups.  If the upper group uses
one metabolic cycle more, then the mutational change  of the parameter
$g^{\ell}$ to enhance the cycle is in favor for the upper group, while
the change to reduce it may be in favor for the lower group.
Indeed all numerical results
support the existence of such parameters.
This dependence of growth rate on the genotypes leads to the
genetic separation of
the two groups, as long as there is competition for survival,
to keep the population numbers limited.

The genetic separation is often accompanied by
a second process, the amplification of the
phenotypic difference by the genotypic difference.  In the
situation of Fig.1c,  as a
parameter $G$ increases, a phenotype $P$ (i.e., a characteristic quantity
for the phenotype) tends to increase
for the upper group, and to decrease (or to remain the same) for the lower group.

It should be noted that this second stage is {\em always} observed in
our model simulation when the phenotypic differentiation at the first stage
occurred. As a simple
illustration, assume that the use of metabolic processes is different
between the two groups. If the upper group uses one metabolic cycle more,
then the mutational change of the parameter $g^{m}$ (e.g., enzyme catalytic
activity) to enhance
the cycle is in favor for the upper group, while the change to reduce it
may be in favor for the lower group. Indeed, all the numerical results carried out
so far support that there always exist such parameters.
This dependence of growth on genotypes leads to genetic separation of the
two groups.

{\bf Stage-III Genetic fixation}

After the separation of two groups has progressed,
each phenotype (and genotype) starts to be preserved by the offspring,
in contrast to the situation at the first stage.
However, up to the second stage, the two groups with different phenotypes cannot
exist in isolation by itself.  When isolated, offspring with the
phenotype of the other group starts to appear.
The two groups coexist depending on each other (see Fig.1d).

Only at this third stage, each group starts to exist by its own.
Even if one group of units is isolated, no offspring with the phenotype of the
other group appears.  Now the two groups exist on their own.
Such a fixation of phenotypes is possible through the evolution of
genotypes (parameters).  In other words, the differentiation is fixed into the genes
(parameters).  Now each group exists as an independent `species',
separated both genetically and phenotypically.
The initial phenotypic change introduced by
interaction is fixed to genes, and the `speciation process' is completed.

At the second stage, the separation is not fixed rigidly. 
Units selected from one group at this
earlier stage again start to show phenotypic differentiation, followed by
genotypic separation, as demonstrated by several simulations.
After some generations, one of the differentiated groups recovers the
geno- and phenotype that had existed before the transplant experiment.
This is in strong contrast with the third stage.

\section{Remarks on the speciation by interaction-induced phenotypic differentiation}

\subsection{Dynamic consolidation to genotypes}

At the third stage, two groups with distinct genotypes and phenotypes
are formed, each of which has one-to-one mapping from
genotype to phenotype.  This stage now is regarded as speciation
(In the next section we will show that this separation satisfies 
hybrid sterility in sexual reproduction, and is appropriate to be called speciation).
When we look at the present process only by observing
initial population distribution (in Fig.1a) and the
final population distribution (in Fig.1d), without information on
the intermediate stages given by Fig. 1b) and 1c), one might think that 
the genes split into
two groups by mutations and as a result, two
phenotype groups are formed, since there is only a flow from
genotype to phenotype.  As we know the intermediate stages, however, 
we can conclude that this simple picture does not hold here.
Here phenotype differentiation  drives the genetic separation,
in spite of the flow only from genotype to phenotype.
Phenotype differentiation is consolidated to genotype,
and then the offspring take the same phenotype as their ancestor.

\subsection{Robust speciation}

Note that the speciation process of ours occurs under strong interaction.
At the second stage, these two groups 
form symbiotic relationship.  As a result, the speciation is
robust in the following sense. If one group is eliminated externally,
or extinct accidentally at the first or second stage, 
the remaining group forms the other phenotype group again,
and then the genetic differentiation is started again.
The speciation process here is robust against perturbations.

\subsection{Co-evolution of differentiated groups}

Note that each of the two groups forms a niche for the other
group's survival mutually, and each of the groups is specialized 
in this created niche.
For example, some chemicals secreted out by one group are used as
resources for the other, and vice versa.

Hence the evolution of two groups are mutually related.
At the first and second stages of the evolution,
the speed for reproduction is not so much different between the two groups.
Indeed, at these stages, the reproduction of each group is strongly dependent
on the other group, and the
`fitness' as a reproduction speed of each group by itself alone
cannot be defined.  At the stage-II, the reproduction of each group is balanced
through the interaction, so that one group cannot dominate in the
population (see Fig.2).

\subsection{Phenotype differentiation is necessary and sufficient for the
sympatric speciation in our theory}

\underline{sufficient}

If  phenotypic differentiation at the stage 1 occurs in our model, then the
genetic differentiation of the later stages {\em always} follows, in spite of
the random mutation process included.  How long it takes
to reach the third stage 
can depend on the mutation rate, but the speciation process
itself does not depend on the mutation rate.  However small the mutation rate
may be, the speciation (genetic fixation) always occurs.

Once the initial parameters of the
model are chosen, it is already determined whether the
interaction-induced phenotype differentiation
will occur or not. If it occurs, then always the genetic differentiation follows.

\underline{necessary}

On the other hand, in our setting,
if the interaction-induced differentiation does not
exist initially, there is no later genetic diversification process. 
If the initial parameters characterizing nonlinear internal dynamics
or the coupling parameters characterizing interaction 
are small, no phenotypic differentiation occurs.  Also, the larger
the resource per individual is, the smaller the effective interaction is.
Then, phenotypic differentiation does not occur.
In these cases, even if we take a large mutation rate, there
does not appear differentiation into distinct genetic
groups, although the distribution of genes (parameters) is broader.
Or, we have also made several simulations starting
from a population of units with widely distributed parameters (i.e.,
genotypes).  However, unless the phenotypic separation into
distinct groups is formed, the genetic differentiation does not follow. (Fig.4a and b)
Only if the phenotype differentiation occurs, the genetic 
differentiation follows(Fig.4c). 

For some other models with many variables and parameters, the phenotypes
are often distributed broadly, but continuously without making
distinct groups.  In this case again, there does not appear distinct genetic groups,
through  the mutations, although the genotypes are broadly distributed.
(see Fig.5).

\section{Post-mating Isolation}

The speciation process is defined both by genetic differentiation and
by reproductive isolation (Dobzhansky 1937).  Although the evolution through
the stages I-III leads to genetically isolated reproductive units, one might
still say that it should not be called `speciation'  unless the process
shows isolated reproductive groups under the sexual recombination.
In fact, it is not trivial if the present process works
with sexual recombination, since the genotypes from parents are mixed by each
recombination.  To check this problem,
we have considered some models so that the sexual recombination occurs
to mix genes.  To be specific, the reproduction occurs when two
individuals $i_1$ and $i_2$ satisfy the maturity
condition, and then the two genotypes are
mixed.  As an example we have produced two offspring $j=j_1$ and $j_2$,
from the individuals $i_1$ and $i_2$ as

\begin{equation}
g^{\ell }(j)=g^{\ell }(i_1)r^{\ell }_j+ g^{\ell }(i_2)(1-r^{\ell }_j) +\delta
\end{equation}
with a random number $0<r^{\ell }_j<1$ 
to mix the parents' genotypes (see also appendix \S 11.2).

In spite of this strong mixing of genotype parameters,
the  two distinct groups are again formed.
Of course, the mating between the
two groups can produce an individual with the parameters in the middle of the two groups.
When parameters of an individual take
intermediate values between those of the two groups, 
at whatever phenotypes it can take,
the reproduction takes much longer time
than those of the two groups. Before the reproduction condition is satisfied, the
individual has a higher probability to be removed by death.
As the separation process to the two groups further progresses, an individual
with intermediate parameter values never reaches the
condition for the reproduction before it dies.

This post-mating isolation process is demonstrated clearly by measuring
the average offspring number of individuals over given parameter (genotype) ranges and
over some time span.  An example of this
average offspring number is plotted in Fig.3, with the
progress of the speciation process.  As the two groups
with distinct values of parameters are formed,
the average offspring number of an individual having
the parameter between those of the two groups
starts to decrease.  Soon the number goes to zero, implying that the
hybrid between the two groups is sterile.

In this sense, sterility (or low
reproduction) of the hybrid appears as a {\bf  result, without any assumption
on mating preference}.  
Now genetic differentiation and reproductive isolation are satisfied.
Hence it is proper to call
the process through the stages I-III as speciation.

\section{Evolution of mating preference}

So far we have not assumed any
preference in mating choice.  Hence, a sterile
hybrid continues to be born.
Then it is natural to expect that some kind of mating
preference evolves to reduce the probability to produce a sterile hybrid. 
Here we  study how mating preference evolves as a
result of post-mating isolation.

As a simple example, it is straightforward to
include loci for mating preference parameters.
We assume another set of genetic parameters that controls
the mating behavior.
For example,
each individual $i$ has a set of mating threshold parameters
$(\rho^1(i), \rho^2(i), \cdots \rho^k(i))$,
corresponding to the phenotype $(X^1(i),X^2(i),\cdots , X^k(i))$.  
If $\rho^{\ell}(i_1)> X^{\ell}(i_2)$ for some $\ell$, the individual 
$i_1$ denies the mating with $i_2$ even if $i_1$ and $i_2$ satisfy
the maturity condition.  In simulation with a model with $\{ \rho^m(i)\}$,
we choose a pair of individuals that salsify the maturity condition, and check
if one does not deny the other.  Only if neither denies the mating
with the other, the mating occurs to produce offspring, when
the genes from parents are mixed in the same way as
as in the previous section.
If these conditions are not satisfied, the individuals $i_1$ and $i_2$
wait for the next step to find a partner again (see also appendix \S 11.3).

Here the set of $\{\rho^m \}$ is regarded as a set of (genetic) parameters, and
changes by mutation and recombination.  The mutation is given by addition
of a random value to $\{ \rho^m \}$.
Initially all of $\{\rho^m\}$ (for $m=1,\cdots ,k$) are smaller than
the minimal value of $(X^1(i),X^2(i),\cdots , X^k(i))$,
so that any mating preference doe not exist. 
If some $\rho^{\ell}(i)$ gets larger than some of $X^{\ell}(i')$,
there appears mating preference.

An example of numerical results is given in Fig.5, where
the change of phenotype $X^{m}$
and some of the parameters $g^{j}$, are plotted.
Here, by the phenotype differentiation, one group (to be called `up' group) has
a large $X^m$ value for some $m=\ell$ and almost null values
for some other $m=\ell '$.  Hence,
sufficiently large positive $\rho^{\ell '}$
gives a candidate for mating preference.  

Right after the formation of two
genetically distinct groups that follows the phenotype separation,
one of the mating threshold parameters ($\rho^1(i_1)$) starts to increase for one group.
In the example of the figure, `up' group has
phenotype with (large $X^1$, small $X^2$) and the other (`down')  group
with (small $X^1$, large $X^2$).  There the
`up' group starts to increase $\rho^1(i_{up})$, and
$\rho^1(i_{up}) > X^1(i_{down})$ is satisfied for an individual $i_{down}$ of the `down' group.
Now the mating between the two groups
is no more allowed, and the mating occurs only within each group.
The mating preference thus evolved prohibits inter-species
mating producing sterile hybrid.

Note that the two groups do not simultaneously establish the mating preference.
In some case, only one group has positive $\rho^{\ell}$, which is enough for 
the establishment of mating preference, while in some other cases
one group has positive $\rho^{1}$, and the other has positive $\rho^{2}$,
where the mating preference is more rigidly established.

Although the evolution of mating preference here is a direct consequence
of the post-mating isolation, it is interesting to note that the 
coexistence of the two species is further stabilized with the establishment of
mating preference.  Without this establishment, there are some cases that one of the
species disappears due to the fluctuation after very long time in the simulation.
With the establishment, the two species coexist much longer ( at least within
our time of numerical simulation).

\section{Formation of allele-allele correlation}

In diploid, there are two alleles, and two alleles
do not equally contribute to the phenotype.  For example,
often only one allele contributes the control of the phenotype.
If by recombination, the loci from two alleles are randomly mixed,
then the correlation between genotype and phenotype achieved
by the mechanism so far discussed might be destroyed.
Indeed, this problem was pointed out by Felsenstein (1981) as 
one difficulty for sympatric speciation.

Of course, this problem is resolved, if genotypes from two alleles
establish high correlation.  To check if this correlation is generated,
we have extended our model to have two alleles, and examined
if the two alleles become  correlated.  Here, we adopted the model studied so far,
and added two alleles further (see also appendix \S 11.4).  In mating, 
the alleles from the parents are randomly shuffled for each locus.
In other words, each organism $i$ has two sets of parameters $\{g^{(+)\ell}(i)\}$ 
and $\{g^{(-)\ell}(i)\}$.  Each $g^{(+)m}(i)$ is inherited
from either $g^{(+)m}$  or $g^{(-)m}$ of one of the parents,
and the other  $g^{(-)m}(i)$ is inherited from  either $g^{(+)m}$  of $g^{(-)m}$ of 
the other parents.  Here parameters at only one of the alleles work as a control 
parameter for the developmental dynamics of phenotype.

We have carried out some simulations of this version of our model (Kaneko, unpublished).
Here again, the speciation proceeds in the same way, through the stages I,II, and
III.  Hence our speciation scenario works well in the presence of alleles.
 
In this model, the genotype-phenotype correspondence achieved at the stage III,
could be destroyed if there were no correlation between two alleles.
Hence we have plotted the correlation between two alleles by showing
two-dimensional pattern $(g^{(+)1}(i),g^{(-)1}(i))$ in Fig.7.
Initially there was no correlation, but through temporal evolution, the
correlation is established.  In other words, the speciation in phenotype is
consolidated to genes, and later is consolidated to the
correlation between two alleles.

\section{Allopatric speciation as a result of sympatric speciation}

As already discussed, our speciation proceeds, starting from
phenotypic differentiation, then to genetic differentiation, and then to
post-mating isolation, and finally to pre-mating isolation
(mating preference).  This ordering might sound strange from commonly
adopted viewpoint, but we have shown that
this ordering is a natural and general consequence of
a system with developmental dynamics with potential plasticity by interaction.

In a biological system, we often tend to assume {\bf causal relationship} between 
two factors, from the  observation of just {\bf correlation} of the
two factors.
For example, when the resident area of two species, which share a common ancestor species,
is spatially separated,
we often guess that the spatial separation is a cause for the speciation.
Indeed, allopatric speciation is often adopted for the explanation of
speciation in nature.

However, in many cases, what we observed in field is just correlation between
spatial separation and speciation.  Which is the cause is not necessarily
proved.  Rather, spatial segregation can be a result of (sympatric)
speciation\footnote{
Consider, for example, the segregation of resident area in city between rich and poor people.
Most of us do not assume that people in `rich area'  are
rich because they live there.  Rather most think that
the spatial separation is a {\em result} of differentiation in wealth, but not 
a {\em cause}.  In the same way, it is sometimes dangerous to assume 
allopatric speciation even if the residence of two species are separated.}.
 
By extending our theory so far, we can show
that spatial separation of two species is resulted from
the sympatric speciation discussed here.
To study this problem, we
have extended our model by allocating to each organism a resident 
position in a two-dimensional space.  Each organism can move around
the space randomly but slowly, while  resources leading  to the competitive interaction
diffuse throughout space much faster.  
If the two organisms that satisfy 
the maturation condition meet in the space (i.e, they are located within a
given distance), then they mate each other to produce offspring.

In this model, we have confirmed that the sympatric speciation first   occurs
through the stages I-III in \S3.  Later these two differentiated 
groups start to be  spatially segregated, as shown in Fig.8.
Now sympatric speciation is shown to be consolidated to
spatial segregation (Kaneko, in preparation).

The spatial segregation here
is observed when the range of interaction is larger than the typical range of
mating.  For example, if mobility of resources causing
competitive interaction is larger than the mobility of organisms,
spatial segregation of symptarically formed species is resulted. 

Instead of spatially local mating process, one can assume
slight gradient of environmental condition, for example, 
as gradient in resources.  In this case again,
sympatric speciation is expected to be later fixed to spatial separation.

To sum up, we have pointed out here the possibility that some of 
speciation that are considered to be allopatric can be a result of sympatric
speciation of our mechanism.  The sympatric speciation is
later consolidated to spatial segregation of organisms.

\section{Comparison with previous theories}

Our theory reviewed here is related with several earlier theories,
but is conceptually different. 
Here we will briefly discuss these points.

\subsection{Frequency-dependent selection}

Since our mechanism crucially depends on the interaction, one might think that
it is a variant of frequency-dependent selection.
The important difference here is that phenotype
may not be uniquely determined by the genotype, even
though the environment (including population of organisms) is given.
In the frequency-dependent selection, 
genetically (and accordingly phenotypically) different groups
interact with each other, and the fitness depends on the
population of each group (Futsuyma, 1986).  At the third stage of our theory,
the condition for this frequency-dependent selection is satisfied,
and the evolution progresses  with the frequency-dependent selection.
However, the important point in our theory lies in the earlier stages
where a single genotype leads to different phenotypes.
Indeed this intrinsic nature of
differentiation is the reason why the speciation process here works
at any (small) mutation rate and also under sexual recombination,
without any other ad hoc assumptions.  

\subsection{Baldwin's effect}

In our theory phenotype change is later consolidated to genotype. Indeed,
genetic `takeover' of phenotype change was also discussed as Baldwin's effect, 
where the displacement of phenotypic character
is fixed to genes.  In the discussion of Baldwin's effect,
the phenotype character is given by
epigenetic landscape (Waddington,1957).  In our case, the phenotype differentiation is
formed through developmental process to generate  different characters
due to the interaction.  Distinct characters are stabilized each other 
through the interaction.  With this interaction dependence, the two groups 
are necessary with each other, and  robust speciation process is resulted.  
Hence, the fixation to genotype in our theory is related with 
Baldwin's effect, but the two are conceptually different.

\subsection{Reinforcement}

Since the separation of two groups with distinct phenotypes is supported by the
interaction, the present speciation mechanism is possible without
supposing any mating preference.  In fact,
the hybrid becomes inferior in the reproduction rate, and
the mating preference based on the discrimination in phenotype
is shown to evolve.   Indeed, a mechanism to amplify the differentiation
by mating preference was searched for as reinforcement since Dobzhansky[1951].
Our theory also gives a plausible basis for
the evolution of mating preference without assuming ad-hoc reinforcement, or
without any presumption on the inferiority in hybrid.

\subsection{Phenotypic plasticity}

Note that our phenotypic differentiation through
development is different from the so called `phenotypic plasticity',
in which a single genotype produces {\bf alternative phenotypes} in {\bf alternative
environments}(Callahan, Pigliucci and Schlichting 1987; Spitze and Sadler 1996;
Weinig 2000).  In contrast, in our case,
distinct phenotypes from a single genotype are formed 
{\bf under the same environment}.  In fact, in our model,
this phenotypic differentiation is necessary to show the later
genetic differentiation.  Without this differentiation,
even if distinct phenotypes appear
for different environments as in `phenotypic plasticity',
genetic differentiation does not follow.
In spite of this difference, it is true that both are concerned with 
flexibility in phenotypes.  Some of phenotypic plasticity so far studied may 
bring about developmental
flexibility of ours, under a different environmental condition.

\subsection{Resource competition}

In our case, competitive interaction is relevant to speciation.
Indeed, coexistence of two (or more) species
after the completion of the speciation is discussed as the resource
competition by Tilman[1976,1981].  Although his theory
gives an explanation for the coexistence, the speciation process
is not discussed, because two individuals with a slight genotypic 
difference can have only a slight difference there.
In our theory, even if the genotypes of two
individuals are the same or only slightly different, their phenotypes
can be of quite different types.
Accordingly, our theory provides a basis for resource competition also.

\section{Relevance of our theory to biological evolution}

General conclusion of our theory is that
sympatric speciation can generally occur under strong interaction,
if the condition for interaction-induced phenotype differentiation is satisfied.
We briefly discuss relevance of our theory to biological
evolution.

\subsection{Tempo in the evolution}

Since the present speciation is triggered by interaction, the
process is not so much random as deterministic. 
Once the interaction among individuals brings about phenotypic
diversification, speciation always proceeds directionally without waiting for a
rare, specific mutation. The evolution in our scenario has a
`deterministic' nature and a fast tempo for speciation, which is different
from a typical `stochastic' view of mutation-driven evolution. 

Some of the phenotypic explosions in the history of evolution
have been recorded as having occurred within short geologic periods.
Following these observations, punctuated equilibrium was
proposed [Gould and Eldegridge 1977].  
Our
speciation scenario possibly gives an interpretation of this punctuated
equilibrium.   It
may have followed the deterministic and fast way of interaction-induced
speciation.

\subsection{An answer why low penetrance is frequent in mutants}

In the process of speciation, the potentiality of a single genotype to 
produce several
phenotypes is consumed and may decline. After the phenotypic 
diversification of a single
genotype, each genotype newly appears by mutation and takes one of the
diversified phenotypes in the population. Thus, the one-to-many
correspondence between the original genotype and phenotypes is consumed. 
Through the present process of speciation,
the potentiality of single genotypes to produce various phenotypes
decreases unless the new genotypes introduce another positive feedback
process to amplify the small difference.

As a result, one may see single genotypes expressing only one 
(or a small number of)
phenotypes in nature. Since most organisms at the present time have gone
through several speciation processes, they may have reduced their
potentiality to produce various phenotypes. According to our theory,
if the organisms have a high potentiality, they will undergo a speciation
process before long and the potentiality will decrease. In other words,
natural organisms tend to lose the potentiality to produce various
phenotypes in the course of evolution. As a reflection on the evolutionary
decline of the potentiality, one can expect that mutant genotypes tend to
have a higher potentiality than the wild-type genotype.
As mentioned in \S 1, the low or incomplete penetrance(Opitz 1981) is known to
often occur in mutants, compared with higher penetrance in a wild type.
Our result is consistent with these observation, since wild types are
in most cases, a consequence of evolution,
where the one-to-one correspondence is recovered, while the mutants can have higher potentiality
to have a loose correspondence.

\subsection{Relevance of developmental plasticity to speciation}

Relationship between development and evolution has been
discussed extensively.
Our theory states the relevance of developmental plasticity
to speciation.
Taking our results and experimental facts into account, one can predict that
organisms emerging as a new species have a high potentiality to
produce a variety in phenotypes.  It is interesting to discuss why
insects, for example, have higher potentiality to speciation from this
viewpoint.  Also examining if living fossils, such as {\it Latimeria
chalumnae}, {\it Limulus} and so forth, have a stable expression of a small number of phenotypes.

In our speciation theory, plasticity is declined through the evolution.
Of course, there should be some occasions when the potentiality is regained, 
so that the evolution continues.  For example, change of environment may
influence
the developmental dynamics to regain loose correspondence, or introduction
of novel degrees of freedom or genes may provide such looseness.  
Endosymbiosis can be one of such causes.
Also, change of the interaction through spatial factor may introduce
novel instability in dynamics, resulting in the loose correspondence.

\subsection{Unified theory for speciation in sexual and asexual (and  unicellular) organisms}

One important point in our theory is that the speciation in
asexual and sexual organisms are explained within the same theory.
Of course, the standard definition of species using hybrid 
sterility is applied only for sexual organisms.  However, it is
true that the asexual organisms, or even bacteria, exhibit discrete
geno- and pheno-types.  
It is suggested that `species',
i.e., discrete types with reproductive isolation,
may exist in asexual organisms (Roberts and Cohan 1995, Holman 1987).
There are also discussions that the potentiality of speciation in
asexual organisms is not lower than the sexual organisms.
In this sense, the present theory sheds a new light to the
problem of speciation in asexual organisms as well.

\subsection{Reversing the order}

According to our theory, sympatric speciation under sexual reproduction
starts first from phenotypic differentiation, and then genetic diversification takes place,
leading to hybrid sterility, and finally the speciation is fixed by mating preference.  
This order may be different from studies most commonly adopted.
Hence, our theory will be verified by confirming this chronic order 
in the field. One difficulty here, however,  lies in
that the process from phenotypic differentiation to the last stage
is rather fast according to our theory.  Still, it may be possible to find 
this order in the field, by first searching for phenotypic
differentiation of organisms with identical genotype
and under the identical environment.  
In this respect, the data of cichlid of Nicaraguan lake may be promising (Wilson, Noack-Kunnmann,
and Meyer 2000), since phenotypic difference corresponding to different
ecological niche is observed even though clear genetic difference is not observed yet.

\subsection{Experimental verification}

Discussion on the mechanism of evolution using past data, however, 
often remains anyone's guess. Most
important in our scenario, in contrast, is its experimental verifiability,
since the process of speciation is rather fast. 
For example, the evolution of {\it E. coli} is observed in the laboratory, as has
been demonstrated by Kashiwagi et al.(1998, 2001) and W.-Z. Xu et al.(1996).
As mentioned in \S 1, phenotypic differentiation of
{\it E. coli} is experimentally  observed when their introduction 
is strong.  Since the
strength of interaction can be controlled by the resources and the
population density, one can check whether or not the evolution in genetic
level is accelerated through interaction-induced phenotypic
diversification (Kashiwagi et al., 2001). 
Examination of the validity of our speciation scenario will give a
first step to such study.

\subsection{Summary: Dynamic Consolidation }

To sum up, we have shown that developmental plasticity
induced by interaction leads to phenotypic differentiation,
which is consolidated to genes.  Thus,  distinct species
with distinct geno- and pheno types are formed.
This leads to hybrid sterility, and later
mating preference evolves.  Further later, this differentiation
can be fixed to correlation in alleles or to spatial
segregation.  How the original differentiation in phenotypes
can be understood as symmetry breaking from a homogeneous state,
in the term of physics,
while successive consolidation of the broken symmetry to
different properties observed at later stages
is more important for biological evolution.
This dynamic process of consolidation is a key issue in
development and evolution (see also (Newman, 2002)).

{\bf acknowledgment}

The author would like to thank Tetsuya Yomo for collaboration
in studies on which the present paper is based.  He would also like to thank
Hiroaki Takagi and Chikara Furusawa for useful discussions, and
Masakazu Shimada, Jin Yoshimura, Masakado Kawata, and Stuart Newman 
for illuminating suggestions.  The present study is 
supported by Grants-in-Aid for Scientific
Research from the Ministry of Education, Culture, Sports,
Science and Technology of Japan (11CE2006).

\section{Appendix: An example of our Model}

\subsection{A coupled map model}

To be specific we consider the following model.

We study a simple abstract model of evolution with an
internal dynamical process for development.  In the model,
each individual $i$
has several (metabolic or other) cyclic processes, and the
state of the $j$-th process
at time $n$ is given by $X^j_n (i)$.  With $k$ such processes,
the state of an individual is given by the set
$(X^1_n (i)$,$X^2_n(i)$, $\cdots ,X^k_n(i))$, which
defines the phenotype.
This set of variables can be regarded as concentrations of
chemicals, rates of metabolic processes, or some
quantity corresponding to a higher function.
The state changes temporally according to
a set of deterministic equations with some parameters.
To be specific, our toy model consists of the following dynamics:

(1) Dynamics of the state:
Here, we split $X_n^j(i)$ into its integer part $R_n^{\ell}(i)$ and the
fractional part $x_{n}^{\ell}(i)=mod[X_{n}^{\ell}(i)]$.
The integer part $R_n^j(i)$ is assumed to give the number of times
the cyclic process has occurred since the individual's birth, while the
fractional part $x_{n}^{\ell}(i)$ gives the phase of oscillation in the
process.  The dynamics of the variables $X_n^j(i)$ consist of a mutual influence
of cyclic processes and interaction with other organisms.
As a simple example, the former is assumed to be given by
$\sum_{m} \frac{a^{\ell,m}}{2}sin(2\pi x_n^m(i))$, while the latter is
given by the competition for resources among the $N_n$ organisms existing
at the moment,
given by $I^{\ell}(i)=p sin(2\pi x_n^{\ell}(i))+\frac{s^{\ell}
-\sum_j p sin2\pi (x_n^{\ell}(j))}{N_n}$.
(The second term comes from the constraint $\sum_i
I^{\ell}(i)=s^{\ell}$, i.e., the condition that $N$ individuals compete
for a given resource $s^{\ell}$ at each time step.  The first term
represents the ability to secure the resource, depending on the state.)
Our toy model is given by

\begin{math}
X_{n+1}^{\ell}(i)=X_n^{\ell}(i)+ \sum_m \frac{a^{\ell m}(i)}{2} sin( 2\pi
x_n^m(i))-\sum_m \frac{a^{m \ell}(i)}{2} sin( 2\pi x_n^{\ell}(i)) \end{math}

\begin{math}
+p sin(2\pi x_n^{\ell} (i))+\frac{s^{\ell} -\sum_j p sin2\pi
(x_n^{\ell}(j))}{N_n}. \end{math}

(2) Growth and Death:
Each individual splits into two when a given condition for growth is satisfied.
Taking into account that the cyclic process corresponds to a metabolic,
genetic or other process that is required for reproduction, we assume
that the unit replicates when the accumulated number of cyclic processes goes
beyond some threshold. Thus, the condition is given by
\begin{math}
\sum_{\ell}X_n^{\ell}(i) \geq Thr
\end{math}
(the maturity condition).
The state $X_n^{\ell}(i)$ is reset to a random value between 0 and 1,
when the corresponding individual splits.  To introduce competition,
individuals are eliminated by a given death condition, as well as by
random removal with a given rate.  As for the former condition,
an individual with $X_n^{\ell}(i) < -10$ (i.e., with a reverse process)
is removed.

(3) Genetic parameter and mutation:
Following the discussion in the text, genes are represented as parameters
in the model,
since the control parameters affect the dynamics of phenotypic variables, but
no direct reverse process exists, as dictated by
the central dogma of molecular biology.  Here, genotypes
are given by a set of parameters $a^{m \ell}(i)$, representing the
relationship between the
two cyclic processes $\ell$ and $m$ ($1 \leq \ell,m \leq k$).
This set of parameters changes slightly through
mutation when offspring is reproduced.
With each division, the parameters $a^{m \ell}$ are changed to
$a^{m \ell}+\delta$ with $\delta$, a random number over
$[-\epsilon,\epsilon]$, with small $\epsilon$, corresponding to the mutation rate.

In the present model, due to the nonlinear nature of the dynamics,
$x_n^{\ell}$ often oscillates in time chaotically or periodically.
Hence it is natural to use 
$X^{\ell}(j)$ including its integer part, as a representation of the phenotype, 
since its integer part represents
the number of cyclic process used for reproduction.

{\bf An alternative model using catalytic reactikon network}

We have also carried out some simulations of a model with reaction network, where $X_t^m(i)$ 
represents the $m$th chemical concentration of an individual $i$.
Each individual gets resources depending on its internal state.
Through the above catalytic reaction process, some products are synthesized
from the resources.  When they are beyond a given threshold, they split to two, 
as given in the model for isologous diversification 
(Kaneko and Yomo, 1994,1997,1999, Furusawa and Kaneko, 1998).
With the increase of the number of individuals,
they compete for resources, while they are removed
randomly to include competition.
Since genes code the catalytic activity of enzymes,
the rate of each reaction in the catalytic network is controlled
by a gene.  Hence,  as a genetic parameter $g^{\ell}$,
the parameter for each reaction rate is adopted.
Through the mutation to this reaction rate, the
speciation process discussed throughout the paper is
also observed (Takagi, Kaneko, Yomo 2000).

\subsection{Sexual Reproduction}

To include sexual recombination, we have extended our 
model so that organisms satisfying the threshold condition
mate to reproduce two offspring.
When they mate, the offspring have parameter values that are
intermediate of those of the parents.  Here, the offspring 
$j=j_1$ and $j_2$ are produced
from the individuals $i_1$ and $i_2$ as

\begin{equation}
a^{\ell m}(j)=a^{\ell m}(i_1)r_j+ a^{\ell m}(i_2)(1-r_j) +\delta
\end{equation}
with a random number $0<r^{\ell }_j<1$ 
to mix the parents' genotypes.

\subsection{Mating Preference}

Here the set of $\{\rho^m \}$ is introduced as a set of (genetic) parameters, and
changes by mutation and recombination.  The mutation is given by addition
of a random value over $[-\delta_{\rho},\delta_{\rho}]$.
Initially $\rho^m \leq 0$ (for $m=1,\cdots ,k$ is set smaller than
the minimal value of $(X^1(i),X^2(i),\cdots , X^k(i))$,
so that any mating preference does not exist. 
If $\rho^m(i)$ gets larger than some of $X^m(i')$ there appears
mating preference.

\subsection{Model with two alleles and random shuffling by mating}

Here we assume that each individual has  two sets of parameters $\{a^{\ell m}(j)\}$,
given by $a^{(+)\ell m}(j)$ and $a^{(-)\ell m}(j)$.  In mating, 
the alleles from the parents are randomly shuffled for each locus $(\ell.m)$.
Each $a^{(+) \ell m}(i)$ is inherited
from either $a^{(+)\ell m(p_1)}$  or $a^{(-)\ell m(p_1)}$ of one of the parents $p_1$,
and the other  $a^{(-)\ell m}(i)(p_2)$ is inherited from either $a^{(+)\ell m(p_2)}$ or 
$a^{(-)\ell m(p_2)}$ of 
the other parents $p_2$.  For the dynamics for $X$,  only $\{a^{(+)\ell m}(i)\}$ 
is used.  The other parameter $a^{(-)\ell m}$ is not used, but can be
used as $\{a^{(+)\ell m}(i_{offspring})\}$ of the offspring after the shuffling.

\subsection{Spatial Model}

To an individual $i$ in the model of \S 11.2 (with sexual reproduction),
spatial position $(x_t(i),y_t(i))$ is assigned.  The individual shows Brownian motion
in the 2-dimensional space, by adding random number over $[-\delta_f,\delta_f]$
to $(x_t(i),y_t(i))$ per each step.  They move within a square of a given suze
with a periodic boundary condition.  If two individuals $i$ and $j$ that satisfy the
maturity condition ($\sum_m X_t^m(i) > Thr$) are within a given distance
$d$, they can reproduce two offspring, which are located  between 
$(x_t(i),y_t(i))$ and $(x_t(j),y_t(j))$.

{\bf References}

\begin{enumerate}



\item
C. A. Beam, R. M. Preparata, M. Himes, D. L. Nanney,
``Ribosomal RNA sequencing of members of the Crypthecodinium cohnii
 (Dinophyceae) species complex; comparison with soluble enzyme studies.
Journal of Eukaryotic Microbiology. 40(5):660-667,
(1993).

\item
Callahan H.S., Pigliucci M., and Schlichting C.D.  1997, 
Developmental phenotypic plasticity: where ecology and evolution meet molecular biology,
Bioessays {\bf 19} 519-525

\item
Coyne J.A., \& Orr H.A.,
`` The evolutionary genetics of speciation",
Phil. Trans. R. Soc. London {\bf B 353} 287-305 (1998)

\item
Darwin C.
{\sl On the Origin of Species by means of natural selection or
        the preservation of favored races in the struggle for life}
        (Murray, London,1859).

\item
Dieckmann U. \& Doebeli M.,
``On the origin of species by sympatric speciation", {\sl Nature} {\bf 400}
354-357 (1999)

\item
Doebeli M.
`` A quantitative genetic competition model for sympatric speciation"
J. Evol. Biol. {\bf 9} 893-909 (1996)

\item
Dobzhansky T., {\sl Genetics and the Origin of Species}
(Columbia Univ. Press. N.Y.) (1937,1951)

\item
Felsenstein J. 1981,
Skepticism towards Santa Rosalia, or why are there so few kinds of animals?,
Evolution {\bf 35} 124-138

\item
Furusawa C. \& Kaneko K.,
``Emergence of Rules in Cell Society: Differentiation, Hierarchy, and Stability"
Bull.Math.Biol.  60; 659-687 (1998)

\item
D. J. Futsuyma, {\sl Evolutionary Biology Second edition}, Sinauer Associates
Inc., Sunderland, Mass (1986).

\item
Gilbert S.F., Opitz J.M., and Raff R.A.  1996, 
Resynthesizing Evolutionary and Developmental Biology,
Developmental Biol. {\bf 173} 357-372

\item
Gould S.J., and Eldredge N.
``Punctuated equilibria: the tempo and mode of evolution  reconsidered", {\sl Paleobiology} {\bf 3},
115-151 (1977)

\item
Holman E., 1987
Recognizability of sexual and asexual species of Rotifers,
Syst. Zool. {\bf 36} 381-386

\item
Holmes L.B., ``Penetrance and expressivity of limb malformations"
{\sl Birth Defects. Orig. Artic. Ser.} {\bf 15}, 321-327 (1979).

\item
D.J. Howard and S.H. Berlocher (eds.)
{\sl Endless Form: Species and Speciation}, Oxford Univ. Press.
(1998)

\item
Kaneko K.
``Clustering, Coding, Switching, Hierarchical Ordering,
and Control in Network of Chaotic Elements"
Physica 41 D, 137-172 (1990)

\item
Kaneko K.
``Relevance of Clustering to Biological Networks",
Physica 75D, 55-73 (1994)

\item
Kaneko K.,
``Coupled Maps with Growth and Death: An Approach to Cell Differentiation",
Physica 103 D; 505-527 (1998)

\item
Kaneko K. \& Yomo T.,
`` Cell Division, Differentiation, and Dynamic
Clustering", Physica 75 D, 89-102 (1994).

\item
Kaneko K. \& Yomo T,
``Isologous Diversification: A Theory of Cell Differentiation ",
Bull.Math.Biol.  59, 139-196 (1997)

\item
Kaneko K. \& Yomo T,
``Isologous Diversification for Robust Development of
Cell Society ", J. Theor. Biol., 199 243-256 (1999)

\item
Kaneko K. \& Yomo T,
``Symbiotic Speciation from a Single Genotype",
Proc. Roy. Soc. B, 267, 2367-2373 (2000)

\item
Kaneko K. \& Yomo T,
``Symbiotic Sympatric Speciation through Interaction-driven Phenotype
Differentiation",
Evol. Ecol. Res. (2002), in press

\item
Kashiwagi A., Kanaya T., Yomo T., Urabe I., ``How small can the difference 
among competitors be for coexistence to occur", 
{\sl Researches on Population Ecology} {\bf 40}, 223 (1998).

\item
Kashiwagi A., Noumachi W., Katsuno M., Alam M.T., Urabe I., and Yomo T.
``Plasticity of Fitness and Diversification Process During an Experimental Molecular 
Evolution",
J. Mol. Evol., (2001) in press

\item
Kawata M. \& Yoshimura J.,
{\sl Speciation by sexual selection in hybridizing populations
without viability selection}, Ev.Ec. Res. 2: (2000) 897-909

\item
Ko E., Yomo T., \&  Urabe I.,
``Dynamic Clustering of bacterial population"
Physica 75D, 81-88 (1994)

\item
Kobayashi C., Suga Y., Yamamoto K., Yomo T., Ogasahara K., Yutani K., and Urabe I
J. Biol. Chem. 272, 23011-6,
``Thermal conversion from low- to high-activity forms of catalase I from
Bacillus stearothermophilus" (1997)

\item
Kondrashov A.S. \& Kondrashov A.F, ``Interactions among quantitative traits
in the course of sympatric speciation", {\sl Nature} {\bf 400}
351-354 (1999)

\item
Lande R., ``Models of speciation by sexual selection on phylogenic traits",
Proc. Natl. Acad. Sci. USA {\bf 78} 3721-3725 (1981)

\item
Maynard-Smith  J.,
``Sympatric Speciation", The American Naturalist
{\bf 100} 637-650 (1966)

\item
Maynard-Smith J. and Szathmary E.,
{\sl The Major Transitions in Evolution} (W.H.Freeman, 1995)

\item
Maynard-Smith J., Burian R., Kauffman S., Alberch P., Campbell J.,
Goodwin B., Lande R., Raup D., and Wolpert L.,  1985
Developmental constraints and evolution, Q. Rev. Biol. {\bf 60} 265-287

\item
S.A. Newman,
{\sl From Physics to Development: The Evolution of Morphogenetic Mechanism}
(to appear in ``Origins of Organismal Form", eds.
G.B. M\"uller and S.A. Newman, MIT Press, Cambridge, 2002)

\item
J.M.Opitz, ``Some comments on penetrance and related subjects",
{\sl Am-J-Med-Genet.} {\bf 8} 265-274 (1981).

\item
Roberts M.S., and Cohan F.M., 1995
Recombination and migration rates in natural populations of
{\sl Bacillus Subtilis} and {\sl Bacillus Mojavensis},
Evolution {\bf 49}: 1081-1094

\item
M.L. Rosenzweig, ``Competitive Speciation",
Biol. J. of Linnean Soc. {\bf 10}, 275-289, (1978)

\item
Schiliewen,U.K., Tautz, D. \& P\"a\"abo S.,
``Sympatric speciation suggested by monophly of crater lake cichilids,
Nature {\bf 368}, 629-632 (1994)

\item
Spitze K. and Sadler T.D.,  1996,
Evolution of a generalist genotype: Multivariate analysis of the adaptiveness
of  phenotypic plasticity, Am. Nat. {\bf 148}, 108-123

\item
Takagi, H., Kaneko K., \& Yomo T.
``Evolution of genetic code through isologous diversification of cellular states",
Artificial Life, 6 (2000) 283-305.

\item
Tilman, D.
``Ecological competition between algae: Experimental confirmation of
resource-based competition theory", {\sl Science} {\bf 192} 463-465 (1976)

\item
Tilman, D.
``Test of resource competition theory using four species of lake Michigan algae",
{\sl Ecology} {\bf 62}, 802-815 (1981)

\item
Turner G.F., \& Burrows M.T.
`` A model for sympatric speciation by sexual selection",
Proc. R. Soc. London {\bf B 260} 287-292 (1995)

\item
Waddington C.H.,
{\sl The Strategy of the Genes},
(George Allen \& Unwin LTD., Bristol, 1957)

\item
Weinig C., 2000,
Plasticity versus Canalization: Population differences in
the timing of shade-avoidance responses
Evolution {\bf 54} 441-451

\item
Wilson A.B., Noack-Kunnmann K., and Meyer A.
``Incipient speciation in sympatric Nicaraguan crater lake cichlid fishes:
sexual selection versus ecological diversification''
Proc. R, Soc. London {\bf B 267} 2133-2141 (2000)

\item
W.-Z. Xu, A. Kashiwagi, T. Yomo, I. Urabe, ``Fate of a mutant
emerging at the initial stage of evolution", {\sl Researches on Population Ecology}
{\bf 38}, 231-237 (1996).

\end{enumerate}

\pagebreak

\begin{figure}
\noindent
\hspace{-.3in}
\epsfig{file=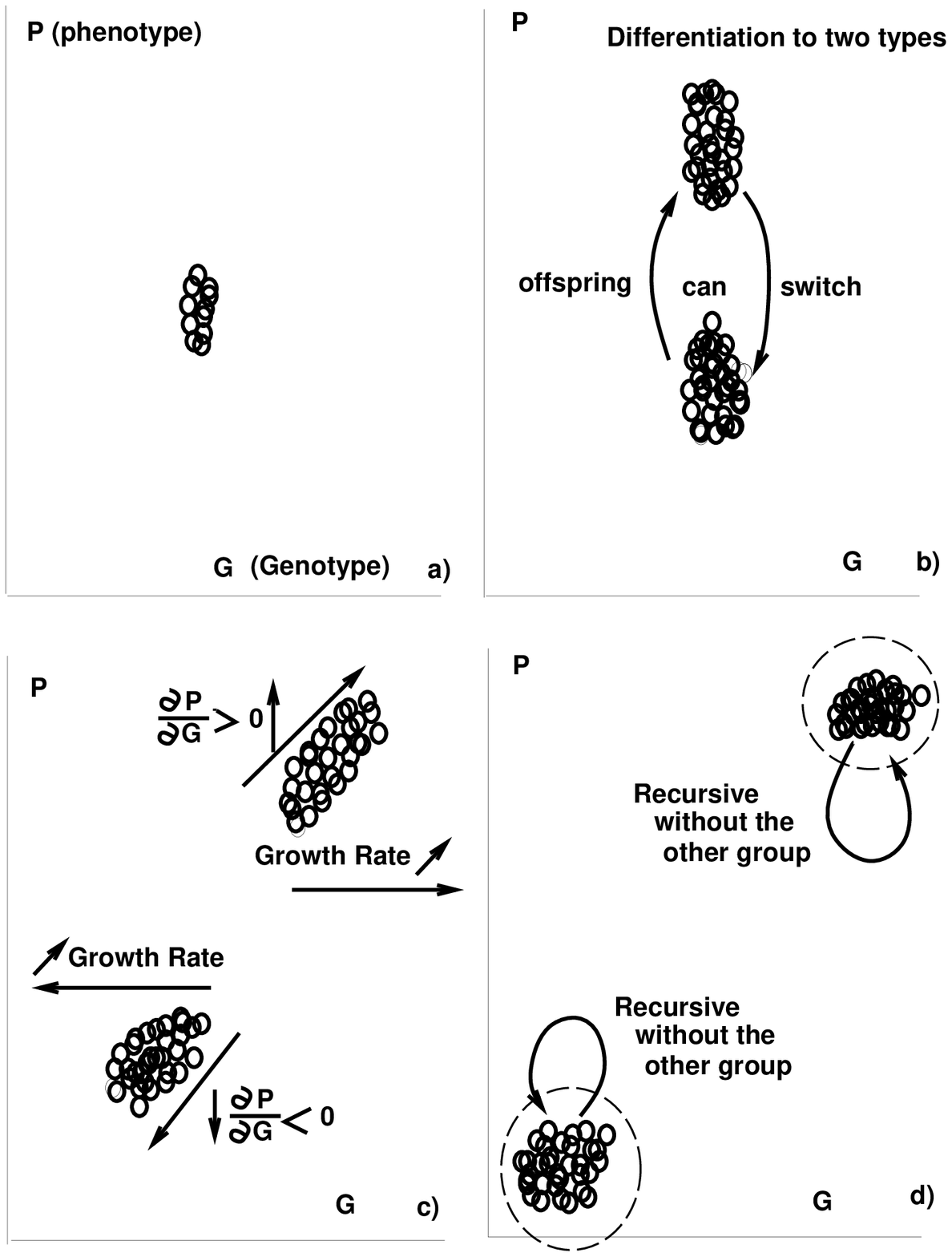,width=.8\textwidth}
\caption{
Schematic representation of the speciation scenario obtained from our
simulation and theory. A pair (phenotype, genotype) is plotted successively
with time: (a) the stage of interaction-induced phenotypic separation (b)
the stage of genotype-phenotype feedback amplification and (c) the stage of
genetic fixation.
(Reproduced from (Kaneko and Yomo 2002)).}
\end{figure}

\begin{figure}
\noindent
\hspace{-.3in}
\epsfig{file=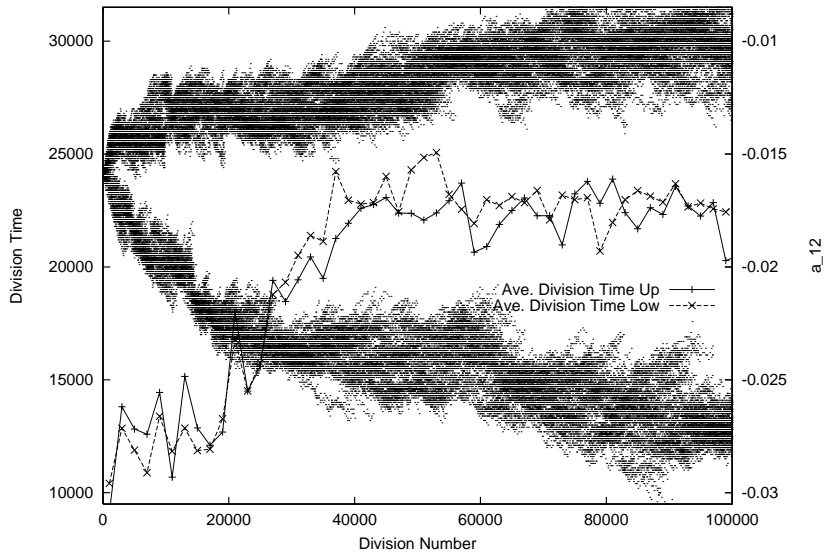,width=.8\textwidth}
\caption{
The evolution of the genotypic parameter.  The parameter $g=a^{12}(i)$ is
plotted as a dot at every division (reproduction) event,
with the abscissa as the division number.
The average time necessary for division (reproduction) is plotted
for the upper and lower groups, where the average
is taken over 2000 division events (6th - 8th generation).
As the two groups are formed around the 2000th division event, the population size
becomes twice the initial, and each division time is also
approximately doubled.  Note that the two average
division speeds of the two groups remain of the same order, even when
the genetic parameter evolves in time.
(Reproduced from (Kaneko and Yomo (2000)).}
\end{figure}

\begin{figure}
\noindent
\hspace{-.3in}
\epsfig{file=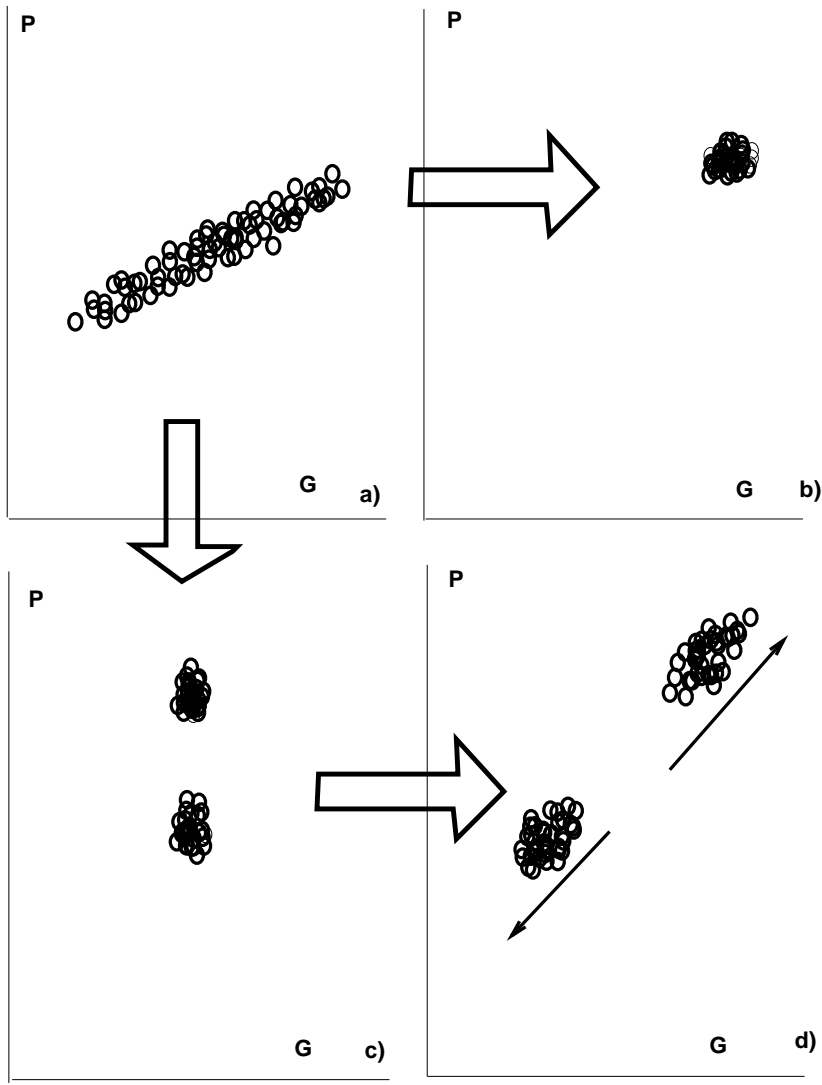,width=.8\textwidth}
\caption{
Schematic representation of the evolution starting from large genetic
variance.  (a) initial genotypic and phenotypic distribution.
(b) without the phenotypic differentiation, no speciation follows.
(c) if phenotypic differentiation occurs, then the speciation follows later (d).}
\end{figure}

\begin{figure}
\noindent
\hspace{-.3in}
\epsfig{file=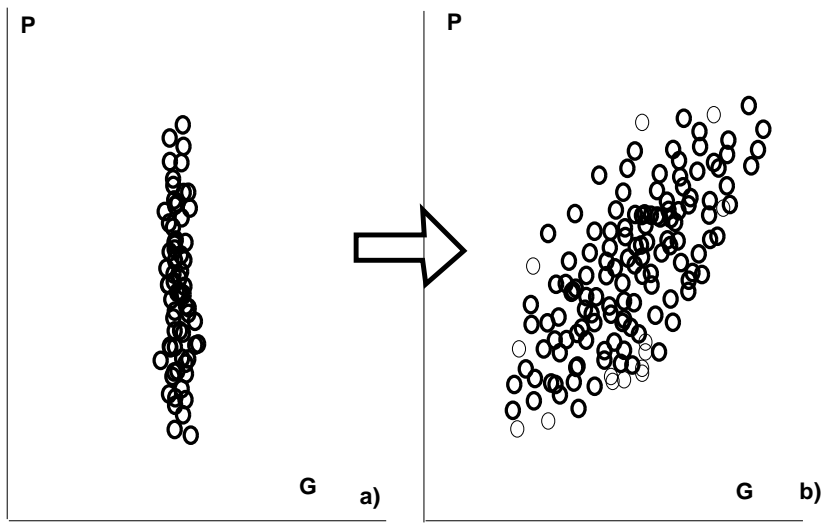,width=.8\textwidth}
\caption{
Schematic representation of the evolution when phenotype is
distributed without clear differentiation to discrete state.
(a) initial distribution of phenotypes and genotypes.
(b) after the evolution, genes are also distributed broadly, but
no speciation follows.
}
\end{figure}

\begin{figure}
\noindent
\hspace{-.3in}
\epsfig{file=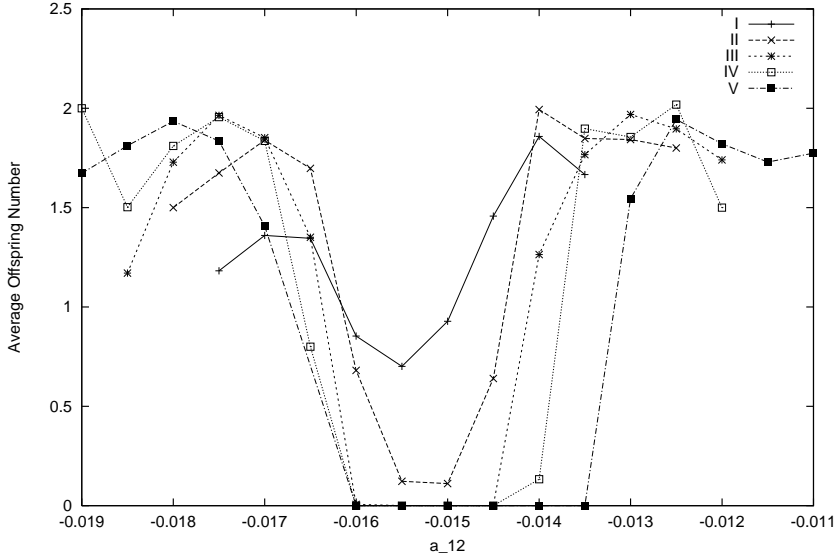,width=.8\textwidth}
\caption{
The average offspring number before death is plotted as a function of
the parameter (genotype), for simulations with sexual recombination.
As an extension to include sexual recombination, we have also studied
a model in which two organisms satisfying the above threshold condition
mate to reproduce two offspring.
When they mate, the offspring have parameter values that are
randomly weighted average of those of the parents, as given in the text.
We have measured the number of offspring for each individual during its lifespan.
By taking a bin width 0.005 for
the genotype parameter $g=a^{12}$, the average offspring number over a
given time span is measured to give a histogram.  The histogram
over the first 7500 divisions (about 20 generations) is plotted by the
solid line (I),
and the histogram for later divisions is overlaid with a different
line, as given by II (over 7500-15000 divisions), III
(1.5-2.25 $\times 10^4$), IV(2.25-3 $\times 10^4$), and V(3.75-4.5 $\times 10^4$).
As shown, a hybrid offspring
will be sterile after some generations.
Here we have used the model of Appendix (\S 11.1,2) and the initial condition as in Fig.1
and imposed recombination, with the parameters
$p_k=1.5/(2\pi)$ and $s^1=s^2=s^3=2$.
In the run, the population fluctuates around 340.
(Reproduced from (Kaneko and Yomo 2000)).
}
\end{figure}

\begin{figure}
\noindent
\epsfig{file=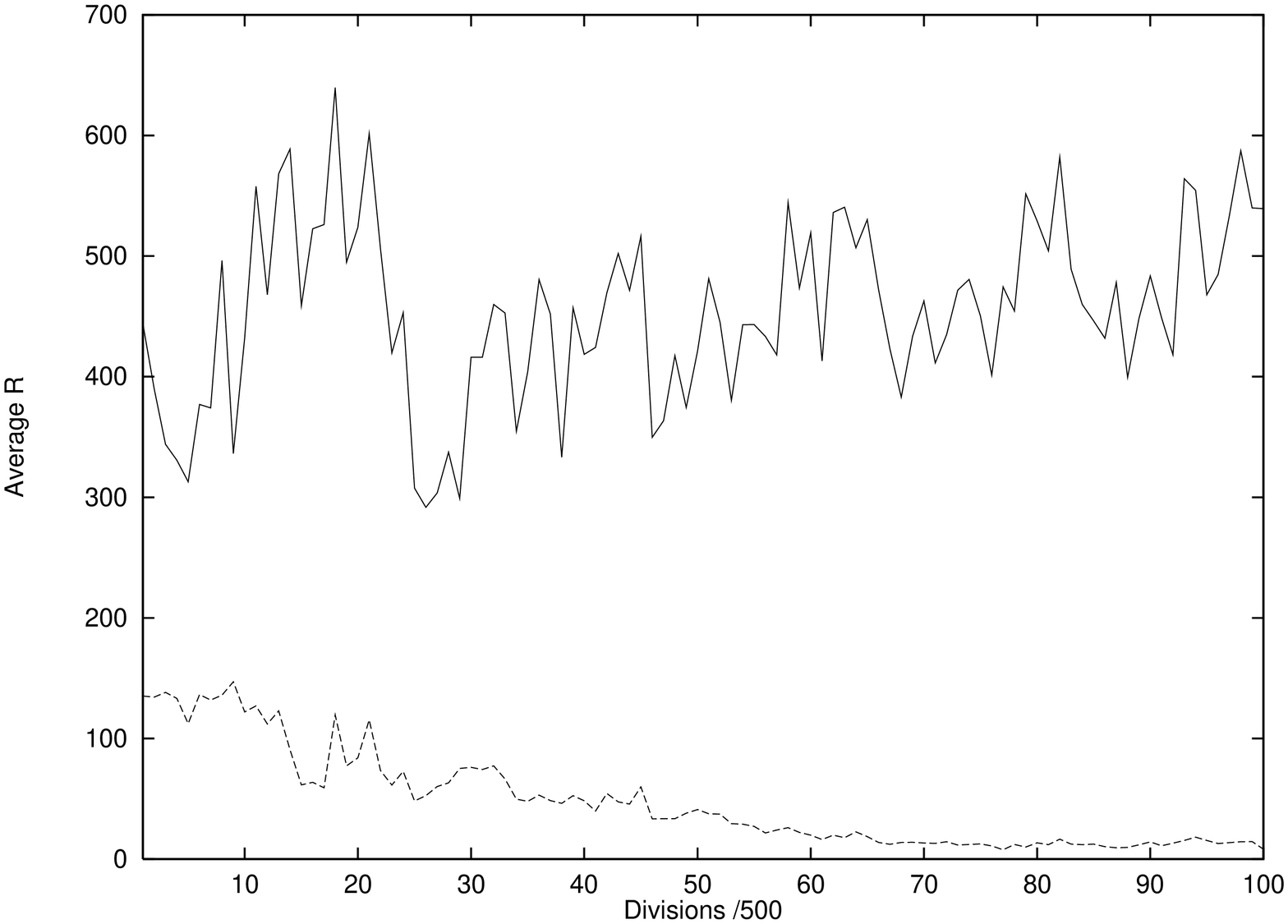,width=.5\textwidth,angle=-90}
\epsfig{file=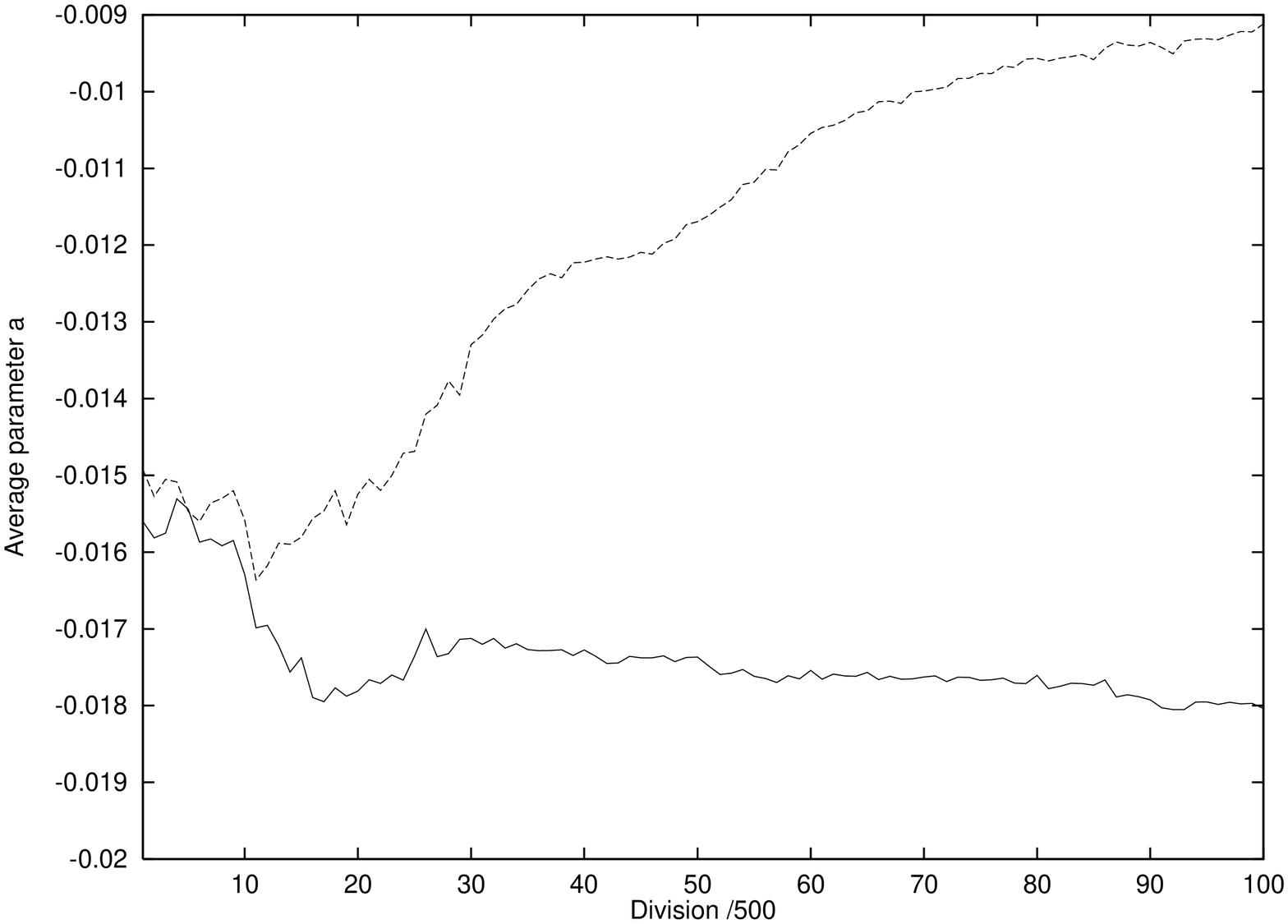,width=.5\textwidth,angle=-90}
\epsfig{file=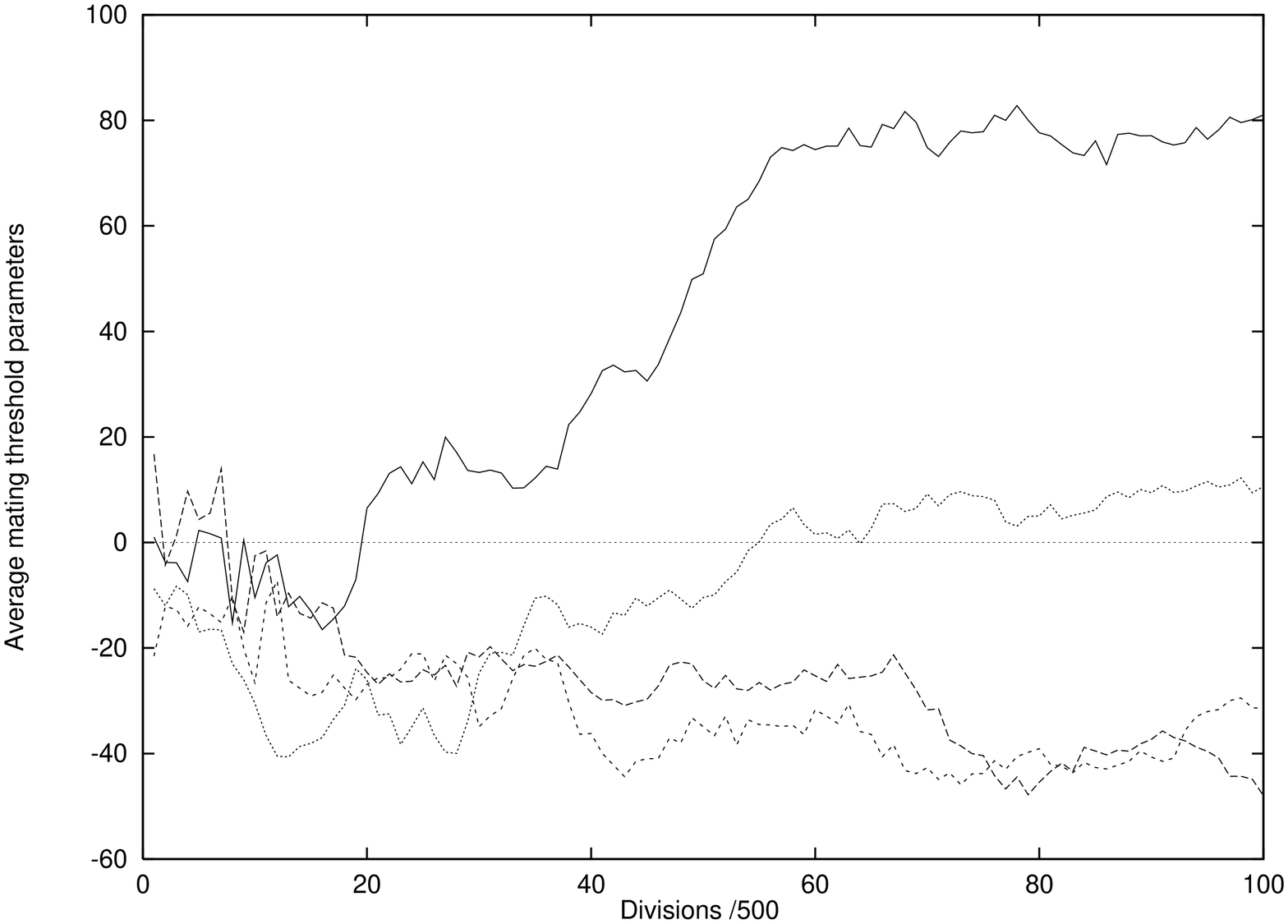,width=.5\textwidth,angle=-90}
\caption{
An example of the speciation process with sexual recombination and the evolution of
mating preference, with the model described in the Appendix (\S 11.1-3).
Here two  groups of distinct phenotype
(large $X^1$, small $X^2$) and (small $X^1$, large $X^2$) are formed at the first few
generation, which we call `up' and `down' groups.
We have measured the average $\overline{X^j}$ at reproduction events, 
$\overline{a^{\ell m}}$, $\overline{\rho^j}$ for
each group per 500 divisions.  ( The population here is roughly 500, and thus the average
is roughly over one generation).  Change of the average $\overline{R^j}$,
$\overline{a^{\ell m}}$, and $\overline{\rho^j}$ are
plotted with divisions (generations). $\delta_{\rho}=20$.
(a) $\overline{X^1}$ (up group ;solid line), $\overline{X^1}$ (down group; broken line),
(b) $\overline{a^{12}}$ (up group ;solid line), $\overline{a^{12}}$ (down group; broken line),
(c) $\overline{\rho^{1}}$ (up group ;solid line), $\overline{\rho^1}$ (down group; broken line),
$\overline{\rho^{2}}$ (up group ;broken line), $\overline{\rho^2}$ (down group; thin broken line),
(Adapted from (Kaneko and Yomo (2002)).
}
\end{figure}

\begin{figure}
\noindent
\epsfig{file=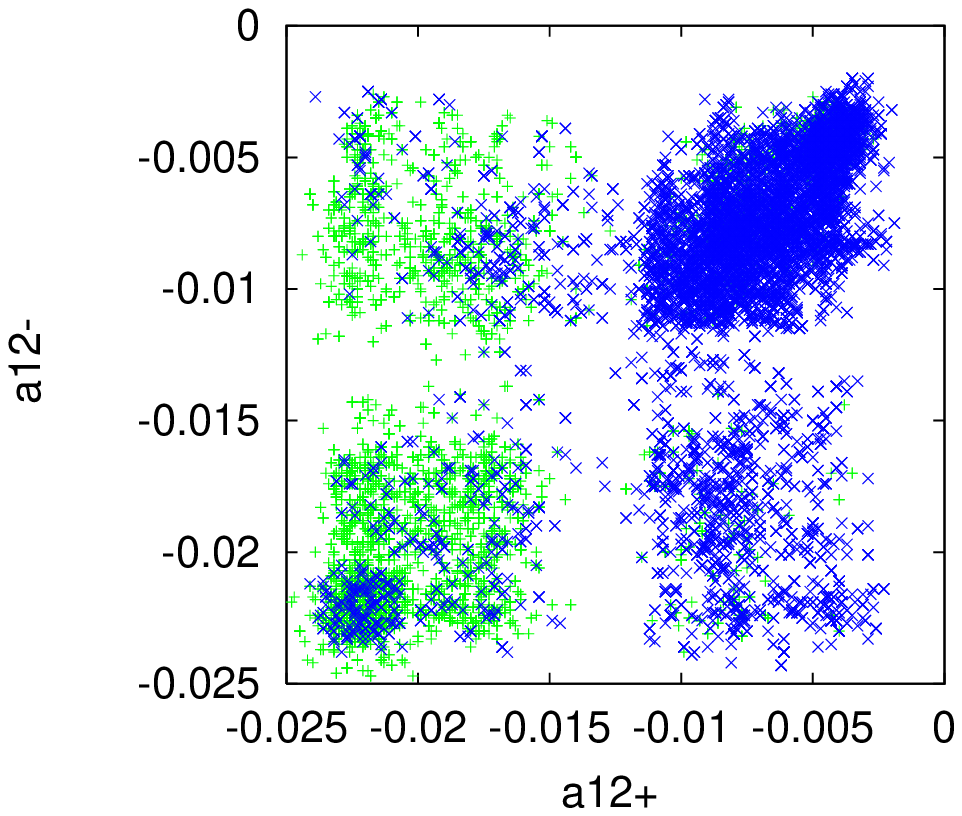,width=.8\textwidth}
\epsfig{file=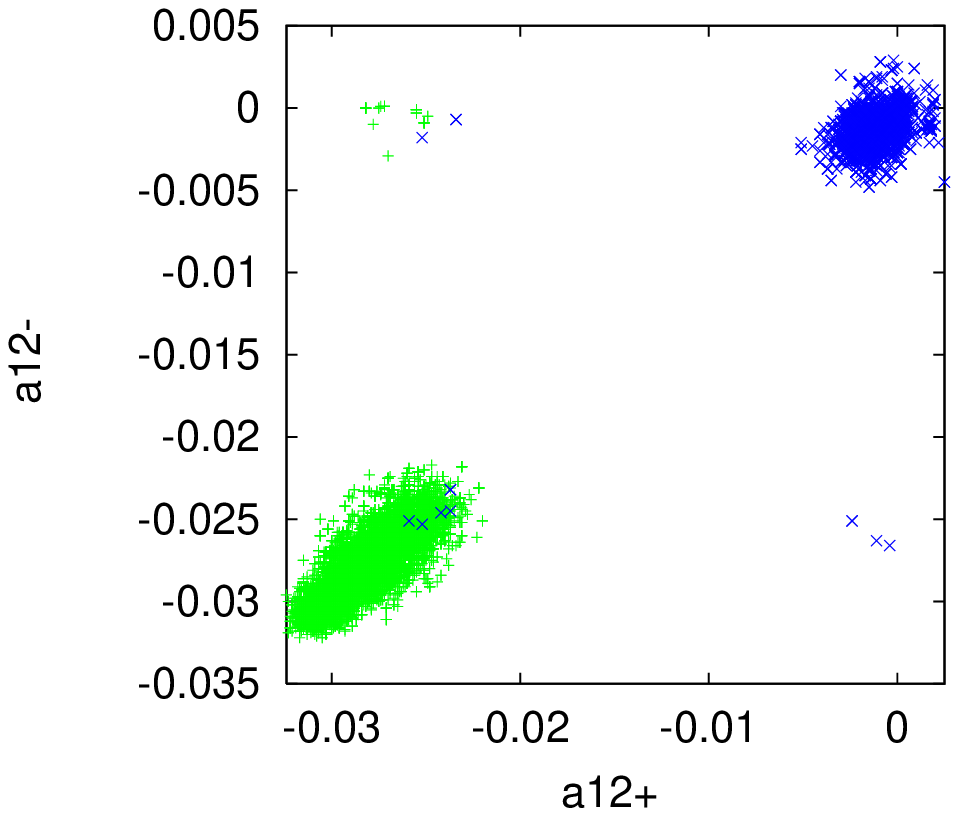,width=.8\textwidth}
\caption{
Correlation in alleles. Using the model in the Appendix (\S 11.1-4), 
the plot of $(a^{(+)12}(i),a^{(-)12}(i))$ is plotted at every reproduction event,
with $p=1.6$, $s_1=s_2=s_3=4$, $\delta_{\rho}=20$.  Initially genotype parameters
$a^{ij}$ is set at $-0.01/(2\pi)$.  Two alternate groups with distinct phenotypes
are plotted with alternate colors.
(a) Plots at the division from 20000 th to 36000th.  Here phenotypic differentiation 
already occurred but the genetic separation is not completed.
(b) Plots at the division at a much later stage (from 50000 th to 63000 th) where 
genetic separation already occurred.
}
\end{figure}

\begin{figure}
\noindent
\hspace{-.3in}
\epsfig{file=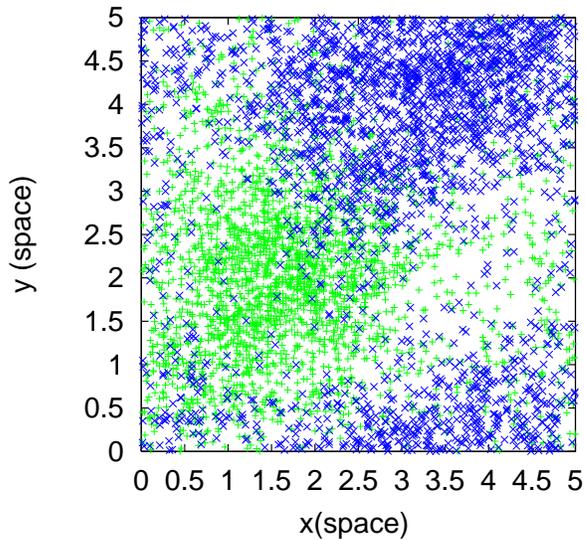,width=.8\textwidth}
\caption{Spatial separation of two species, observed in the model of the Appendix (\S 11.5) .
Model in spatially local mating. Each unit moves the position with the
Brownian motion, given as the random number over $[-\delta_f,\delta_f]=[-0.0025,0.0025]$, while
units are located within 
2-dimensional square of the size 5x5, with periodic boundary condition.
Mating is possible if two units satisfying maturing condition
are located within the distance 0.25.
The parameters are set as  $p=1.6$, $s_1=s_2=s_3=2$, while
initial genotype parameters $a^{ij}$ is set at $-0.01/(2\pi)$.  
Position of the units at every division is plotted,
at each division event from 45000 to 50000.
Two alternate groups with distinct pheno- and geno- types are plotted
with alternate colors.
}
\end{figure}

\end{document}